\documentclass[twocolumn,english,aps,prb,twocolum,superscriptaddress,natbib,bibnotes,amsmath,amssymb,floatfix,groupedaddress,footinbib]{revtex4-2}

\usepackage[markup=nocolor, authormarkupposition=left]{changes} 

\usepackage{soul}
\usepackage[utf8]{inputenc}
\usepackage[english]{babel}
\usepackage{amsmath,amsfonts,amssymb}
\usepackage[T1]{fontenc}
\usepackage{url}

\usepackage{amsmath}
\usepackage{siunitx}
\usepackage[version=4]{mhchem}
\usepackage{amsfonts}
\usepackage{amssymb}

\usepackage[colorlinks=true,citecolor=blue,linkcolor=magenta]{hyperref}

\usepackage{epstopdf}
\usepackage{graphicx}
\graphicspath{{./Figures/}}

\usepackage{changes}

\begin{document}

\title{Autonomous multifunctional image processing via programmable multimode lasing}

\author{Jiawei Wu$^{1,2,3}$, 
        Yue Yin$^{1,2,3}$, 
        Jianqi Hu$^{4}$,       
        Hao Wang$^{5^\ast}$,
        Xing Fu$^{1,2,3^\ast}$,
        and Qiang Liu$^{1,2,3^\ast}$}
\affiliation{
$^1$Department of Precision Instrument, Tsinghua University, Beijing, China.\\
$^2$Key Laboratory of Photonic Control Technology, Ministry of Education, Tsinghua University, Beijing, China.\\
$^3$State Key Laboratory of Precision Space-time Information Sensing Technology, Beijing, China.\\
$^4$Department of Electrical and Computer Engineering, The University of Hong Kong, Hong Kong, China.\\
$^5$Department of Electrical Engineering, City University of Hong Kong, Hong Kong, China.\\
}

\maketitle

\noindent\textbf{\noindent
Optical image processing offers a promising pathway to overcome the latency and energy limitations of conventional electronic image processors. However, existing approaches based on passive photonic devices are often constrained by signal attenuation, lack of nonlinearity, fixed functionality, and heavy training overhead. Here, we introduce a programmable image processor based on a highly multimode degenerate cavity laser (DCL), shifting the computational framework from passive extracavity transformation to active intracavity evolution. By manipulating the intracavity loss distribution and exploiting the nonlinear lasing dynamics, we map computational tasks to the spontaneous mode selection within the DCL, realizing image processing directly at the source. We experimentally demonstrate that by simply altering the encoding scheme of input images, the platform can be flexibly reconfigured for multifunctional tasks, including high-fidelity edge detection and robust image denoising. The computation proceeds autonomously without dataset training, featuring an intrinsically low latency ($\sim$90~$\mu\mathrm{s}$) with $\mathcal{O}(1)$ complexity. Furthermore, the high intensity sustained within the resonator enables intracavity nonlinear upconversion of the processed image. This work extends the frontiers of laser applications, providing a compelling candidate for next-generation optical processors.
}

\section*{Introduction} 

\begin{figure*}[!htp]
  \centering{
  \includegraphics[width = 0.95\linewidth]{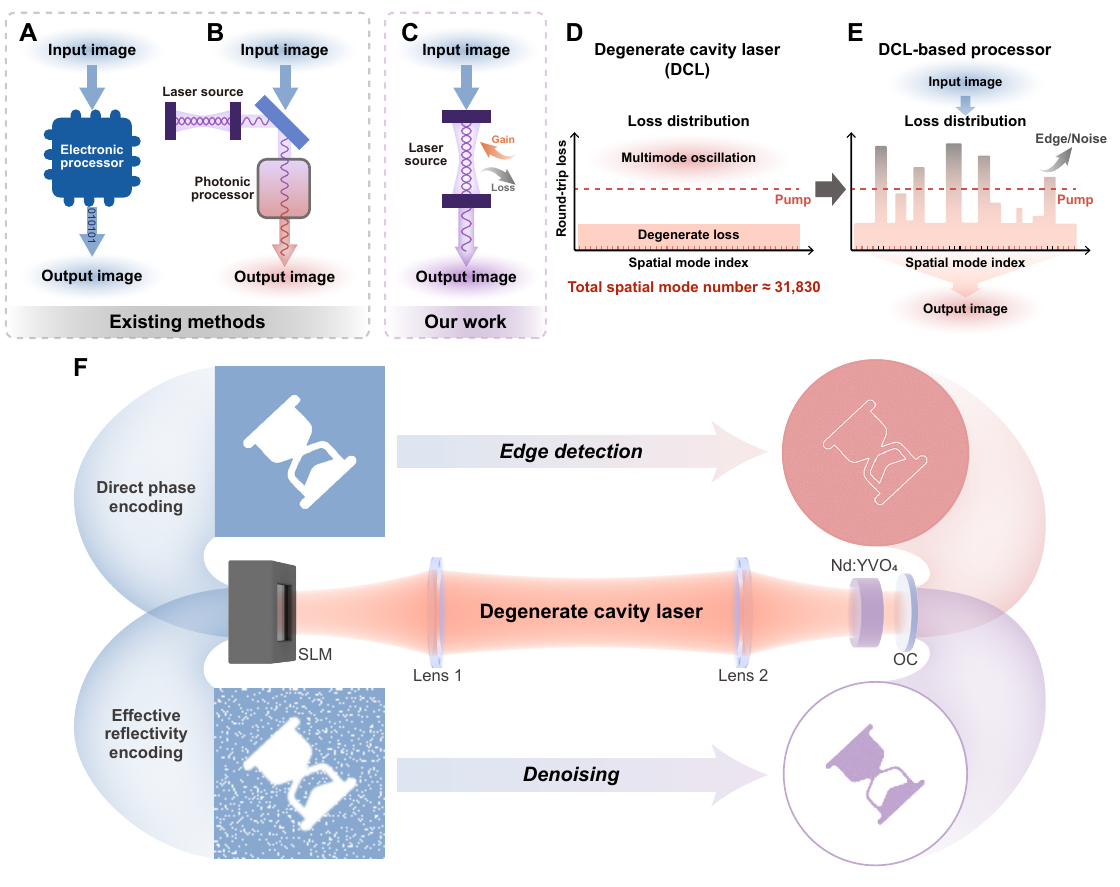}
  } 
    \caption{\noindent\textbf{Multifunctional image processing based on a multimode degenerate cavity laser (DCL).}
    (\textbf{A}--\textbf{C}) Different image processing frameworks: (A) digital electronic processing, (B) passive extracavity optical processing, and (C) the proposed active intracavity optical processing. 
    (\textbf{D} and \textbf{E}) Mechanism of image processing based on intracavity mode selection. (D) An unmodulated DCL supports massive transverse modes with near-identical round-trip losses. (E) By introducing the input image via intracavity modulation, modal degeneracy is broken. The output image is constituted by the selected lasing modes.
    (\textbf{F}) Schematic of the DCL-based image processor. The laser system comprises a 4$f$ system (Lens 1 and Lens 2), a solid-state gain medium (Nd:YVO$_4$), an output coupler (OC), and a reflective phase-only spatial light modulator (SLM) acting as a programmable cavity mirror. By switching the SLM encoding strategies (i.e., direct phase encoding and effective reflectivity encoding), the system seamlessly transitions between two distinct tasks (i.e., edge detection and image denoising) without further hardware modifications.}
    \label{f1}
\end{figure*}

\noindent{Image} processing serves as the cornerstone of numerous modern information technologies, ranging from machine vision and autonomous navigation to computational microscopy and biomedical diagnostics\cite{mennel2020ultrafast,gornet2024automated,ma2024pretraining,li2024integrated}. These advanced applications rely heavily on fundamental operations, including edge detection and image denoising. While conventional digital approaches---spanning from analytical algorithms\cite{canny1986computational,rudin1992nonlinear,dabov2007BM3D} to data-driven neural networks\cite{xie2015HED,zhang2017DnCNN}---benefit from the high reconfigurability and mature infrastructure of electronic processors (Fig.~\ref{f1}A), they are increasingly constrained by the saturation of Moore's law and the von Neumann bottleneck\cite{mehonic2022brain,schuman2022opportunities}. These limitations have motivated growing interest in optical image processing that exploits the massive parallelism, high bandwidth, and low dissipation inherent to wave propagation\cite{wetzstein2020inference,shastri2021photonics,mcmahon2023physics}.

Current optical image processing architectures typically encode images onto a light field and perform computations using extracavity photonic structures, such as Fourier optical filters\cite{furhapter2005spiral}, metasurfaces\cite{zhou2020flat,zhou2019optical,yu2026double,swartz2024broadband,liu2024edge,tanriover2023metasurface,fu2022ultracompact}, photonic crystals\cite{guo2018photonic}, and diffractive neural networks\cite{icsil2024all,zhou2025all} (Fig.~\ref{f1}B). Over recent decades, these platforms have enabled extensive demonstrations of optical edge detection and image denoising, alongside continuous improvements in bandwidth, miniaturization, and scalability. Despite this progress, several critical challenges remain. First, passive photonic processors inevitably attenuate signal amplitude, which severely degrades the signal-to-noise ratio\cite{zhang2025integrated,moralis2024perfect}. This energy budget constraint becomes more critical in scenarios where mid-infrared signals must be upconverted to the visible band for efficient detection\cite{qiu2018spiral,zhao2023high,fang2024wide,xomalis2021detecting}. Second, most existing architectures are restricted to linear wave transformations, inherently limiting their processing capacity. Executing nonlinear operations, which are essential for sufficient noise truncation and contrast enhancement\cite{zhang2025integrated,huang2019programmable}, typically requires either nonlinear materials or electrical-domain participation\cite{chen2024ultra,zhou2025all,wang2023image,williamson2019reprogrammable,song2024low}, thus compromising the optical processing advantages. Third, many modern data-driven platforms, such as diffractive neural networks\cite{icsil2024all,zhou2025all,chang2025few}, incur substantial computational overhead associated with dataset collection and training. Finally, most optical processors remain inherently single-purpose, with their functionality fixed upon fabrication to perform a specific task. Although multifunctional platforms have recently emerged, transitioning between tasks often requires cumbersome hardware-level adjustments, such as temperature tuning\cite{cotrufo2024reconfigurable,chamoli2025reconfigurable,yang2025nonlocal} or illumination wavelength shifting\cite{chamoli2025nonlocal}, which restrict flexible, real-time reconfiguration.

To overcome the inherent constraints of passive optical processors, an alternative paradigm lies in the principle of physical optimization\cite{vadlamani2020physics}, where the dynamics of a physical system are engineered to encode and solve a computational problem. Among such physical systems, lasers, which exhibit complex dynamics and spontaneous self-organization, provide a compelling platform for nontrivial signal processing. One existing strategy leverages nonlinear lasing dynamics for implicit feature extraction, with the laser serving as a physical reservoir or nonlinear hidden layer\cite{brunner2013parallel,porte2021complete,skalli2025model}. This data-driven scheme targets neural network applications. Another approach directly maps computational problems onto the intracavity evolution, such that the steady-state oscillation explicitly represents the solution. Degenerate cavity lasers (DCLs)\cite{arnaud1969degenerate,nixon2013observing,cao2019complex,arwas2022anyonic}, which support a vast number of transverse modes with flexible tunability, are well-suited for the latter framework. Recent studies have spatially discretized the multimode intracavity field into coupled laser channels to simulate many-body systems\cite{pando2024synchronization} and solve optimization problems\cite{gershenzon2020exact}. In parallel, the evolution of the spatially continuous transverse field in DCLs has been successfully exploited for wavefront shaping\cite{nixon2013real} and phase retrieval\cite{tradonsky2019rapid}.

Here, building upon this mechanism, we extend laser-based computing to optical image processing. Rather than relying on extracavity passive components, we elevate the laser resonator from a mere illumination source to the core computing module, realizing image processing directly at the source (Fig.~\ref{f1}C). By incorporating a phase-only spatial light modulator (SLM) into a DCL, we effectively shape the intracavity loss distribution according to the input image. Governed by the interplay between gain and loss, the resonant field autonomously evolves into the desired processed output without any dataset training. Benefiting from the low-coherence multimode lasing field and the inherent nonlinear thresholding mechanism, our laser solver achieves high-fidelity edge detection and robust noise truncation. Switching between the two tasks is realized simply by altering the programmable image encoding strategy, requiring no further hardware modifications. Our platform leverages active stimulated amplification and resonant enhancement to sustain an intense intracavity field, which is sufficient for frequency upconversion. Moreover, because the computational latency is determined solely by the timescale of lasing stabilization ($\sim$90~$\mu\mathrm{s}$) and is independent of the image dimension ($N \times N$), the DCL processor exhibits an $\mathcal{O}(1)$ computational complexity. This fundamentally circumvents the $\mathcal{O}(N^2)$ scaling bottleneck that plagues standard digital algorithms, allowing the system to process dynamic inputs in near real-time. Collectively, this work establishes multimode laser physics as a computing primitive, offering a reconfigurable, high-speed route for versatile optical information processing.

\section*{Results} 
\noindent\textbf{Principle of programmable loss distribution.} As illustrated in Fig.~\ref{f1}F, our optical image processor is built upon a DCL mainly comprising a 4$f$ system (Lens 1 and Lens 2), a solid-state gain medium (Nd:YVO$_4$), an output coupler (OC), and a reflective phase-only SLM serving as a programmable cavity mirror (see Materials and methods). Due to the self-imaging property of the 4$f$ system architecture, the DCL supports a massive number of transverse modes ($\sim$31,830 modes in our system; see Supplementary Note~2) with near-identical round-trip losses, facilitating highly multimode oscillation (Fig.~\ref{f1}D). By loading the input image onto the SLM with a tailored encoding strategy, we deliberately break this modal degeneracy, thereby creating a customized, non-uniform loss distribution across the transverse modes (Fig.~\ref{f1}E). Specifically, distinct local features of the input image, such as object regions, background regions, edges, and noise, are mapped onto corresponding spatial mode subsets with varying round-trip losses. Under a uniform optical pump, the DCL operates as a driven-dissipative system in which the steady-state emission is governed by the nonlinear interplay between gain and loss. Modes associated with edges or noise are suppressed, while the surviving lasing modes constitute the output optical field carrying the processed image. Therefore, the image processing task is effectively mapped to the spontaneous mode selection within the laser cavity.

Crucially, this laser solver is reconfigurable and multifunctional: simply changing the SLM encoding strategy enables seamless switching between two distinct image processing tasks without any hardware modifications (Fig.~\ref{f1}F). For edge detection, we apply a direct phase encoding scheme such that image gradients are translated to phase gradients. These phase discontinuities introduce pronounced local diffraction losses, suppressing edge-associated modal components and thus delineating edge features in the output. For image denoising, we employ an effective reflectivity encoding strategy. This imposes different diffraction losses on the image background, connected object regions, and isolated noise pixels, facilitating the selective amplification of the target object while suppressing both background and noise. In both scenarios, the optical image processing is completed in a single-shot manner without prior training.

\vspace{0.1cm}

\begin{figure*}[!htp]
  \centering{
  \includegraphics[width = 0.9\linewidth]{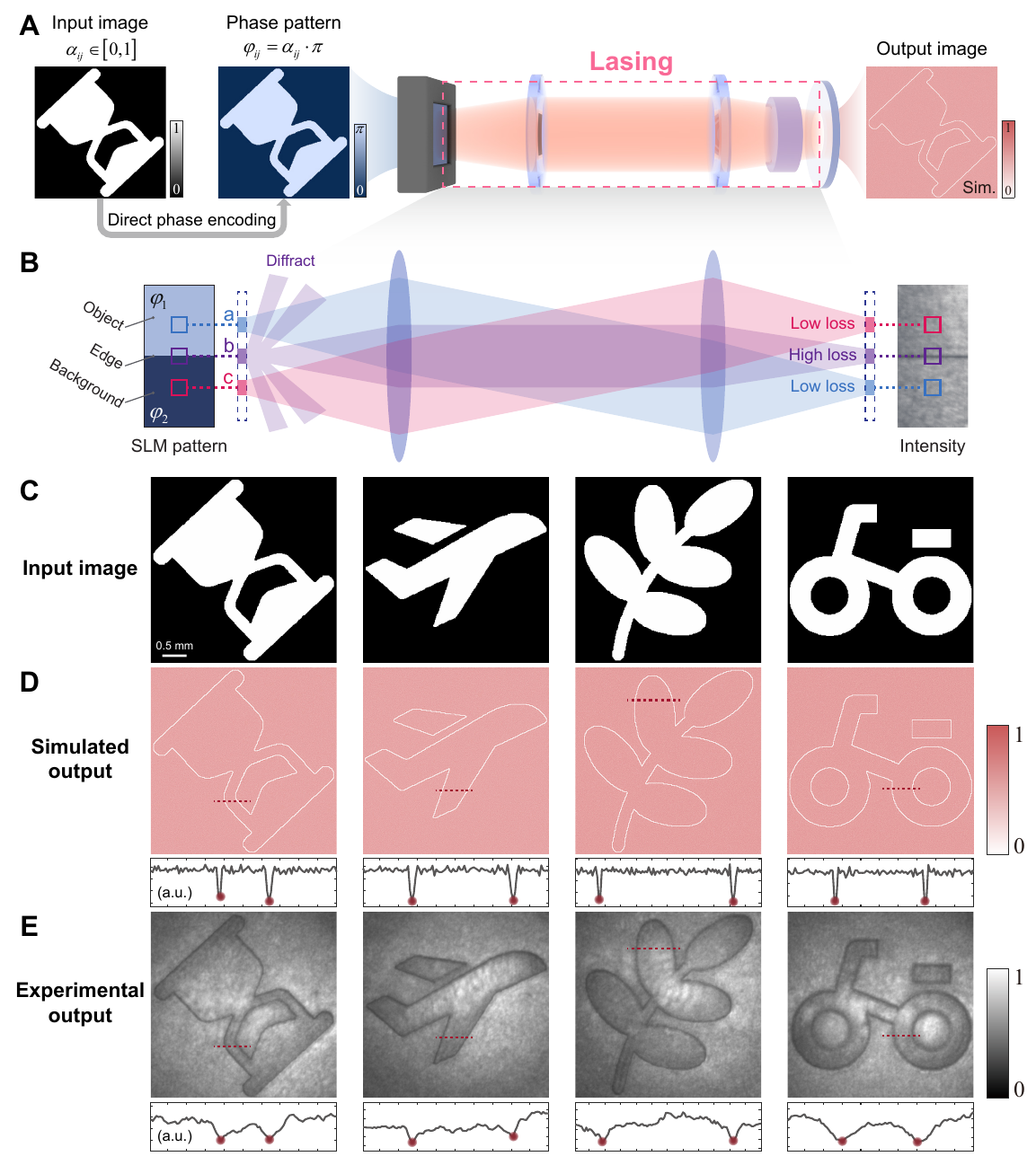}
  } 
    \caption{\noindent\textbf{Edge detection via direct phase encoding.}
    (\textbf{A}) Conceptual workflow. From left to right: the encoding strategy applied to the SLM, the DCL processor, and the simulated processed output.
    (\textbf{B}) Schematic of the physical mechanism.
    (\textbf{C}--\textbf{E}) Representative edge detection results for four complex binary inputs. Row (C) shows the binarized input images. Rows (D) and (E) display the corresponding steady-state outputs obtained from Fox--Li simulations and experiments, respectively. Insets: cross-sectional intensity profiles along the red dashed lines; red circles denote the edge locations identified by a valley-finding algorithm.
  }
  \label{f2}
\end{figure*}

\noindent\textbf{Edge detection.} The ``direct phase encoding'' method for edge detection is illustrated in Fig.~\ref{f2}A. For an input image with pixel values $\alpha_{ij}\in[0,1]$ (where $i$ and $j$ denote the row and column indices), the SLM phase is programmed as $\varphi_{ij}=\pi\cdot\alpha_{ij}$. This linear mapping encodes gray-level variations of the image into spatial phase gradients on the SLM plane. Figure~\ref{f2}B visualizes how the imposed phase pattern sculpts the loss distribution to detect edges. Owing to DCL's highly multimode nature, the intracavity transverse field is a superposition of parallel spatial modes, each sampling a distinct local region of the SLM (e.g., modes a--c in Fig.~\ref{f2}B). The lasing behavior of each mode is governed by its local gain-loss interplay, with the steady-state laser output formed by all simultaneously oscillating modes. More precisely, in spatially flat regions of the input image (i.e., object and background), the SLM phase is locally uniform; consequently, the corresponding modes (a and c) experience minimal diffraction losses and lase efficiently, producing high output intensity. Conversely, at image edges, the large phase gradients introduce strong local diffraction losses to the associated modes (b). These high-loss, edge-associated modal components are selectively suppressed under the same pump level. Finally, edge locations appear as distinct intensity minima against a bright background.

We first validate the edge detection capability on complex binary images (Fig.~\ref{f2}C). As illustrated in Figs.~\ref{f2}D and E, the experimental outputs show that image edges do appear as pronounced intensity minima, consistent with the Fox--Li simulation results (see Materials and methods). Quantitatively, cross-sectional intensity profiles (along the red dashed lines) reveal clear edge locations, which are automatically identified using a valley-finding algorithm (see Materials and methods). The experimental intensity distributions are less uniform than the simulations, primarily due to the pump inhomogeneity.

\begin{figure*}[!htp]
  \centering{
  \includegraphics[width = 0.95\linewidth]{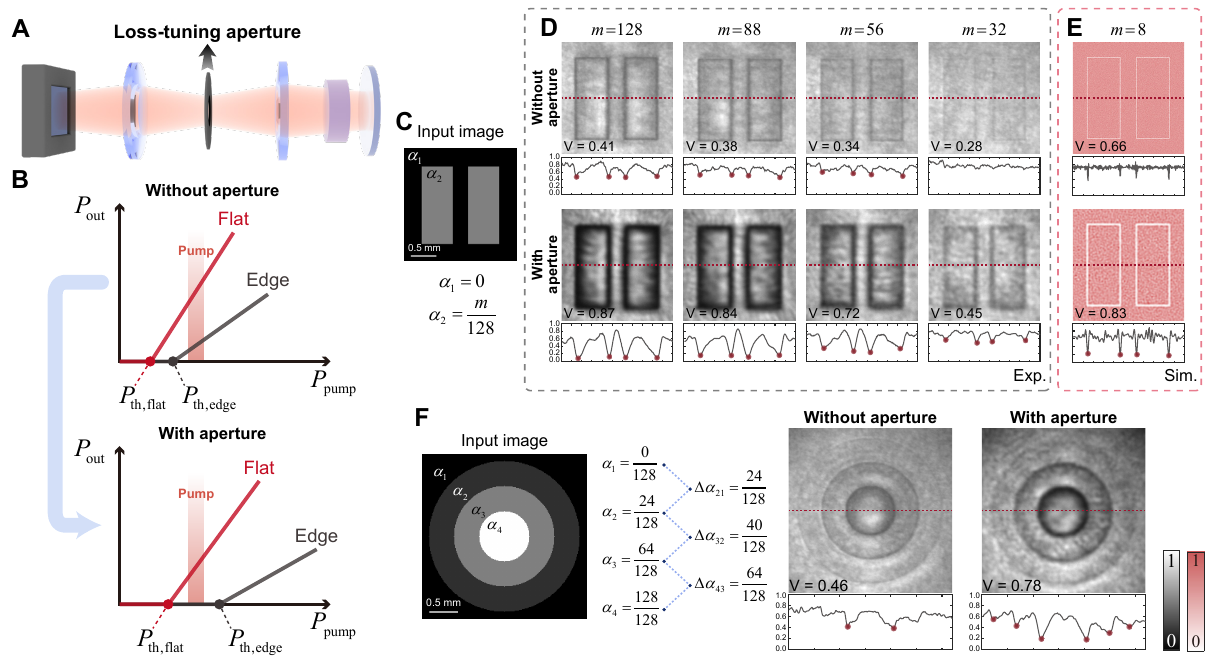}
  } 
    \caption{\noindent\textbf{Edge detection sensitivity to varying gray-level contrast.}
    (\textbf{A}) Schematic of the loss-tuning aperture inserted at the intracavity Fourier plane.
    (\textbf{B}) Schematic illustration of the output power ($P_{\mathrm{out}}$) versus pump power ($P_{\mathrm{pump}}$) for flat-region-associated modes (red curve) and edge-associated modes (gray curve), without (top) and with (bottom) the loss-tuning aperture. The shaded red vertical bar indicates the applied pump power.
    (\textbf{C}) Input pattern featuring a tunable gray-level step height, $\Delta\alpha=m/128$. 
    (\textbf{D}) Experimental outputs without (top) and with (bottom) the loss-tuning aperture for step parameters $m=128, 88, 56,$ and $32$ (from left to right).
    (\textbf{E}) Simulated outputs without (top) and with (bottom) the loss-tuning aperture for step parameter $m=8$.
    (\textbf{F}) Edge detection results for a multi-level grayscale target. Left: input pattern comprising concentric rings with unequal gray-level steps. Right: experimental outputs obtained without and with the loss-tuning aperture, demonstrating enhanced edge visibility across all contrast levels.
  }
  \label{f3}
\end{figure*}

To improve edge detection sensitivity, we manipulate the global loss by introducing a loss-tuning aperture at the intracavity Fourier plane (Fig.~\ref{f3}A; see Supplementary Note~1). As shown in Fig.~\ref{f3}B, edge-associated modes inherently possess a higher pump threshold than flat-region-associated modes due to their higher diffraction losses (see Materials and methods). Without the aperture, the relatively low overall cavity loss allows the pump power to readily exceed both thresholds ($P_{\mathrm{pump}}>P_{\mathrm{th,edge}}>P_{\mathrm{th,flat}}$; Fig.~\ref{f3}B, upper panel). Consequently, edge modes oscillate weakly alongside flat modes, yielding only shallow intensity dips at the image boundaries. In contrast, introducing the Fourier aperture deliberately increases the global diffraction loss, thereby elevating the lasing thresholds for all modes. This shifts the system into a highly sensitive threshold-clipping regime where the pump power falls exactly between the two thresholds ($P_{\mathrm{th,edge}}>P_{\mathrm{pump}}>P_{\mathrm{th,flat}}$; Fig.~\ref{f3}B, lower panel). As a result, flat modes sustain oscillation while edge modes are strictly cut off, creating a sharp contrast between distinct image features. 

We evaluate the enhanced sensitivity using two-level test patterns with a tunable gray-level step $\Delta\alpha=m/128$ ($1\le m \le 128$; Fig.~\ref{f3}C). Without the loss-tuning aperture (Fig.~\ref{f3}D, top row), edge-induced intensity dips gradually vanish as $m$ decreases, leading to a monotonic decline in edge visibility $V$ (see Materials and methods), until reliable detection fails at $m=32$. In contrast, applying a 9-$\mathrm{mm}$-diameter aperture (Fig.~\ref{f3}D, bottom row) significantly deepens these depressions, approximately doubling the visibility and ensuring robust edge detection even at $m=32$. Numerical simulations at an extremely low gray-level step ($m=8$) exhibit excellent consistency with these observations (Fig.~\ref{f3}E). We further extend our evaluation to a multi-level grayscale target with unequal steps (Fig.~\ref{f3}F). In both configurations, image boundaries manifest as dark rings, whose depths scale with the gray-level step. The insertion of the loss-tuning aperture markedly accentuates these edge signatures across all steps, increasing the overall visibility from $V=0.46$ to $V=0.78$. Thus, the DCL processor's applicability is broadened to complex, multi-level grayscale patterns with non-uniform gradients. Notably, this spatial constraint reduces the effective numerical aperture (NA) of the resonator, limiting the number of supported transverse modes ($\sim$2,783 modes in our system; see Supplementary Note~2). However, driven by the lasing dynamics, the intracavity aperture guides the resonant field to reorganize its wavefront, facilitating spontaneous energy redistribution rather than simple energy blocking\cite{cao2019complex}. Consequently, the energy efficiency is largely maintained (see Supplementary Note~8). This property fundamentally distinguishes our method from extracavity passive filtering.

\begin{figure*}[!htp]
  \centering{
  \includegraphics[width = 0.9\linewidth]{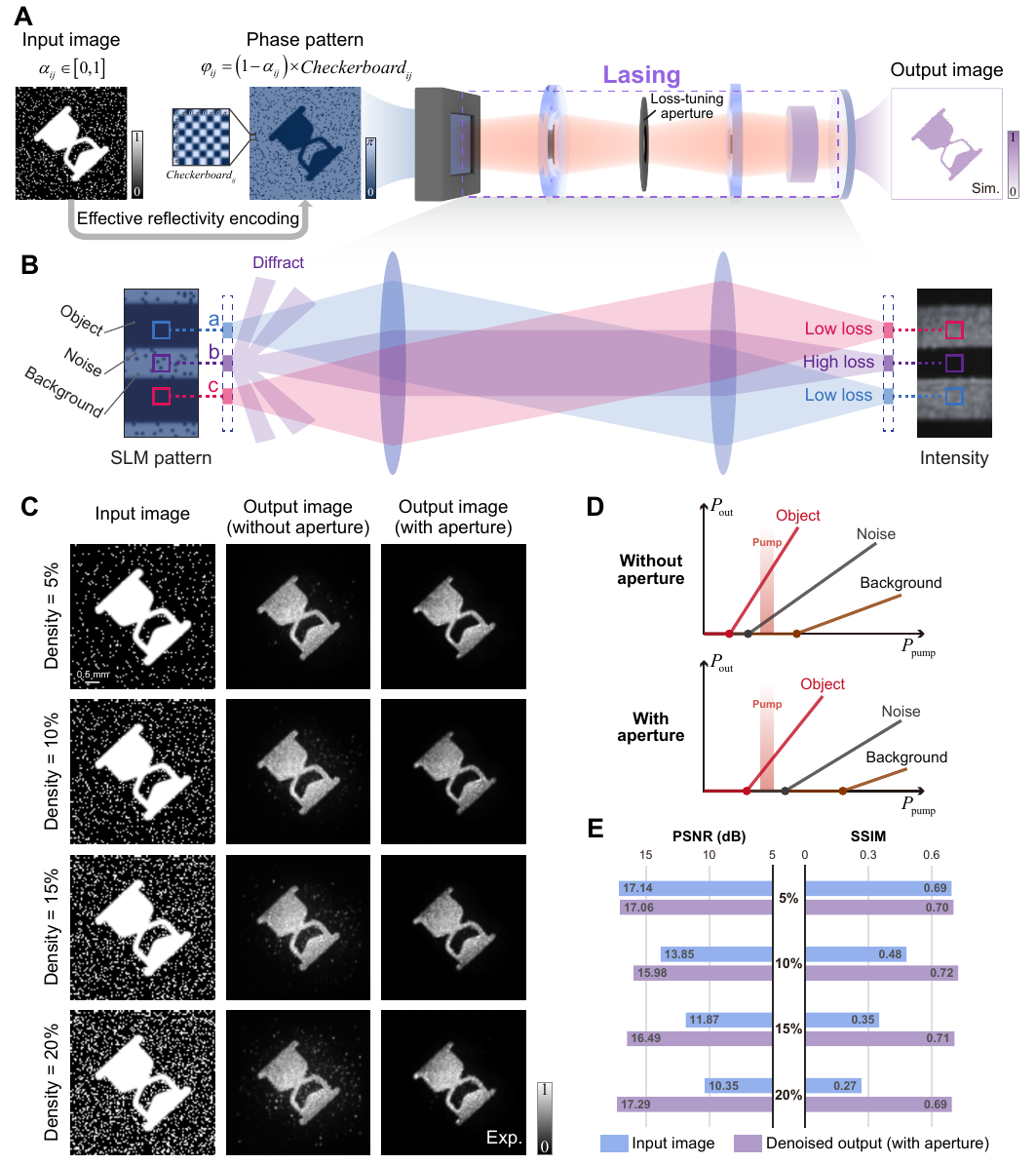}
  } 
    \caption{\noindent\textbf{Image denoising via effective reflectivity encoding.}
    (\textbf{A}) Conceptual workflow. From left to right: the SLM encoding strategy, the DCL processor with a loss-tuning aperture, and the simulated denoised output.
    (\textbf{B}) Schematic of the physical mechanism.
    (\textbf{C}) Experimental results for noise densities from 5\% to 20\% (top to bottom). Left: noisy inputs. Middle and right: outputs obtained without and with the loss-tuning aperture, respectively.
    (\textbf{D}) Schematic illustration of the output power ($P_{\mathrm{out}}$) versus pump power ($P_{\mathrm{pump}}$) for object (red), noise (gray), and background (brown) modes, without (top) and with (bottom) the loss-tuning aperture. The shaded red vertical bar indicates the applied pump power.
    (\textbf{E}) PSNR and SSIM metrics comparing the noisy inputs (blue) and the denoised outputs (purple).
  }
  \label{f4}
\end{figure*} 

\vspace{0.1cm}

\noindent\textbf{Image denoising.} For image denoising, we employ an ``effective reflectivity encoding'' strategy (Fig.~\ref{f4}A). The SLM phase pattern is programmed as the element-wise product of the inverted input image and a binary phase mask ($Checkerboard_{ij}$) with alternating $0$ and $\pi$ phase values: $\varphi_{ij}=(1-\alpha_{ij})\times Checkerboard_{ij}$. Consequently, the dark background ($\alpha_{ij}\approx 0$) is mapped into a checkerboard pattern. This region acts as a low-reflectivity mirror, scattering incident light into large-angle diffraction components that escape the finite cavity\cite{tradonsky2021high}. Conversely, the bright object ($\alpha_{ij}\approx 1$) maps to a spatially uniform phase distribution, acting as a high-reflectivity mirror with minimal diffractive leakage. Crucially, while the bright noise pixels are also encoded with a nearly uniform phase, their smaller spatial extent inherently introduces higher diffraction losses. Through this mechanism, the lasing thresholds associated with distinct image features become clearly differentiated (Fig.~\ref{f4}D). Thus, modes sampling both the background and noise regions are attenuated or clipped (e.g., mode b in Fig.~\ref{f4}B), while only modes within the extended object region (e.g., modes a and c) sustain stable oscillation.

We experimentally evaluate this denoising performance under noise densities ranging from 5\% to 20\% (Fig.~\ref{f4}C; see Supplementary Note~5 for additional results). Without the loss-tuning aperture, the background noise is attenuated, yet residual noise patterns remain visible at higher noise densities (Fig.~\ref{f4}C, middle column). As before, inserting the 9-$\mathrm{mm}$-diameter aperture at the intracavity Fourier plane dramatically improves the performance. The output background becomes nearly spotless, while the object remains clearly resolved even under a heavy noise density of 20\% (Fig.~\ref{f4}C, rightmost column). This improvement arises from the enhanced thresholding mechanism depicted in Fig.~\ref{f4}D: with the loss-tuning aperture, the system enters a regime where noise modes are strictly cut off by the pump threshold, while the object modes are preserved ($P_{\mathrm{th,noise}}>P_{\mathrm{pump}}>P_{\mathrm{th,object}}$). This nonlinear thresholding mechanism for effective noise suppression is systematically verified in Supplementary Notes~6 and 7.

To quantitatively evaluate denoising efficacy, we compute the Peak Signal-to-Noise Ratio (PSNR) and the Structural Similarity Index (SSIM) in Fig.~\ref{f4}E (see Materials and methods). Both metrics indicate that as noise density increases, the raw input image degrades dramatically with severe pixel-wise corruption and structural distortion. However, the denoised output maintains consistently high metric scores, demonstrating substantial recovery of the image. Notably, at the most challenging 20\% noise density, the system improves the PSNR by $\sim$7~$\mathrm{dB}$ and increases the SSIM by 0.42.

\begin{figure*}[!htp]
  \centering{
  \includegraphics[width = 0.9\linewidth]{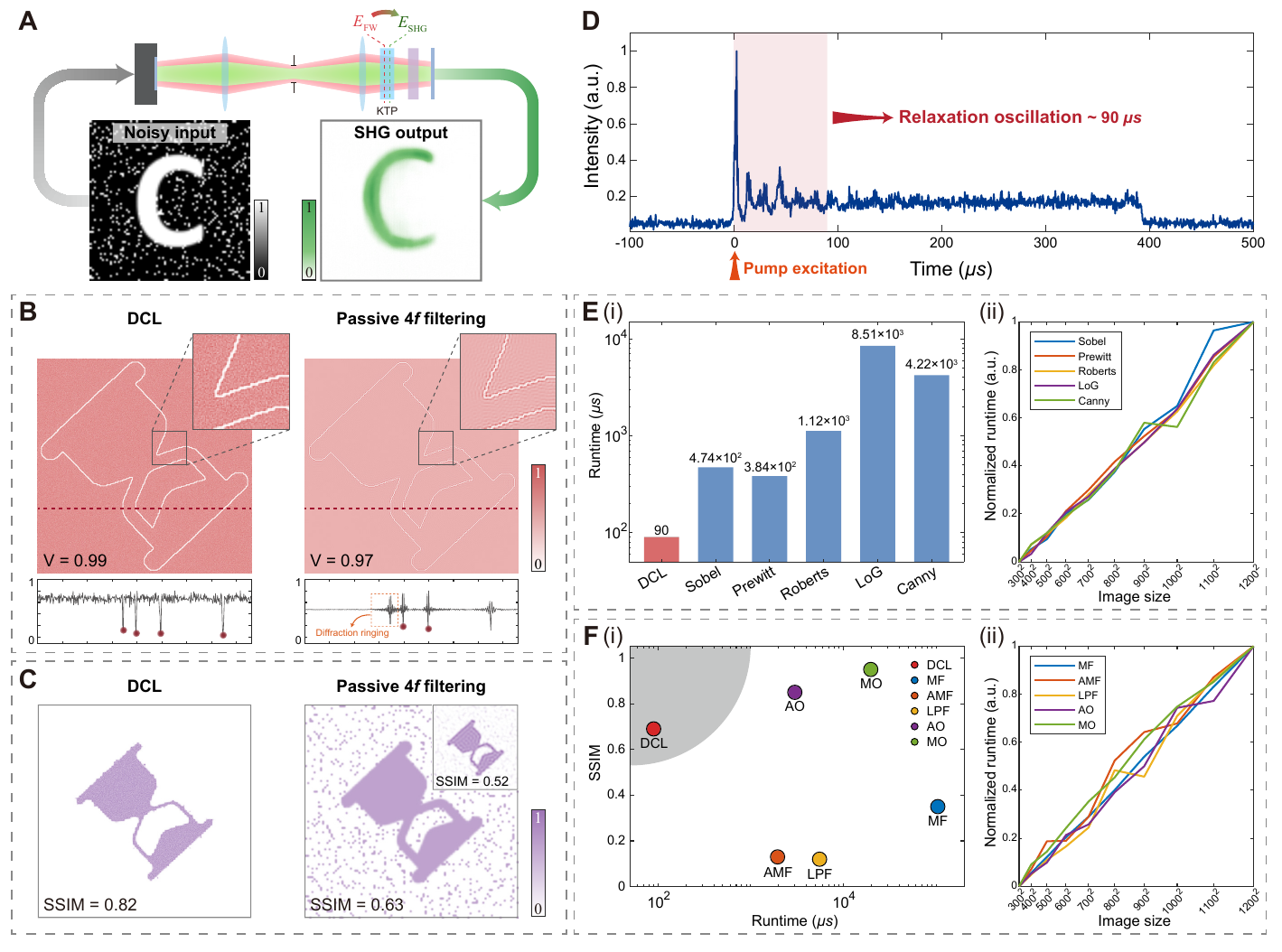}
  } 
    \caption{\noindent\textbf{Benchmarking the computational performance of the DCL-based processor.}
    (\textbf{A}) Experimental demonstration of intracavity second-harmonic image processing, which is unique to the DCL processor. The schematic (top) illustrates the integration of a nonlinear potassium titanyl phosphate (KTP) crystal. For a given noisy input (bottom left), the denoised 1064-$\mathrm{nm}$ fundamental field formed within the cavity is upconverted into a denoised 532-$\mathrm{nm}$ output (bottom right).
    (\textbf{B} and \textbf{C}) Numerical comparison between the DCL and a passive 4$f$ filtering system for (B) edge detection and (C) image denoising.
    (\textbf{D}) Experimentally measured temporal evolution of the DCL's output intensity, revealing a physical latency of $\sim$90~$\mu\mathrm{s}$. The inherent computational complexity of the DCL is $\mathcal{O}(1)$ (independent of image size).
    (\textbf{E}) Benchmarking latency against digital edge detection algorithms. (i) Absolute runtimes for processing a $300 \times 300$ image. (ii) Normalized runtimes as a function of image size ($N \times N$), revealing the quadratic scaling $\mathcal{O}(N^2)$ complexity of digital algorithms.
    (\textbf{F}) Benchmarking performance and latency against digital image denoising algorithms. (i) Joint evaluation of SSIM metrics and absolute runtimes for processing a $400 \times 400$ image. (ii) Normalized runtimes as a function of image size ($N \times N$), revealing the quadratic scaling $\mathcal{O}(N^2)$ complexity of digital approaches.
    MF, median filtering; AMF, arithmetic mean filtering; LPF, low-pass filtering; AO, area opening; MO, morphological opening.
  }
  \label{f5}
\end{figure*} 

\vspace{0.1cm}

\noindent\textbf{Benchmarking the computational performance of the DCL-based processor.} Distinct from passive extracavity optical systems, the active DCL sustains an intense circulating field by leveraging stimulated amplification, resonant enhancement, and intracavity energy redistribution (detailed in Supplementary Note~8). Therefore, nonlinear crystals can be integrated into the cavity to upconvert infrared image signals into the visible spectrum, enabling sensitive detection with mature and cost-effective sensors\cite{zhao2023high,fang2024wide,xomalis2021detecting}. To demonstrate this, we perform an intracavity second-harmonic generation (SHG) experiment. As depicted in Fig.~\ref{f5}A, a nonlinear potassium titanyl phosphate (KTP) crystal is inserted into the DCL, upconverting the fundamental lasing field (1064~$\mathrm{nm}$) to its second harmonic wave (532~$\mathrm{nm}$). When a noisy image is input, we successfully achieve a visible, noise-suppressed output (see Supplementary Note~9 for detailed experimental setup). Remarkably, this nonlinear SHG of the low-coherence field operates in a temporally quasi-continuous-wave regime without strict phase-matching control, reflecting the DCL's high intracavity energy advantage over most passive optical processing systems.

Numerical simulations further highlight the DCL's performance superiority over passive 4$f$ optical filtering (detailed in Supplementary Note~11). For edge detection (Fig.~\ref{f5}B), enabled by the incoherent superposition of massive transverse modes, the DCL's output exhibits a smooth intensity profile. In contrast, the output of the passive 4$f$ system suffers from significant ringing artifacts near the edges due to the high-coherence illumination. These artifacts distort the edge profile, causing our valley-finding algorithm to miss two of the four edges. For image denoising (Fig.~\ref{f5}C), the performance gap is more pronounced. The DCL leverages the lasing threshold to nonlinearly clip the noise modes, yielding a clean output (SSIM = 0.82). Conversely, under the same aperture constraint at the Fourier plane, the passive 4$f$ system can only linearly attenuate noise components, leaving a residual noisy background (SSIM = 0.63). Although reducing the Fourier aperture size visually suppresses some noise (top-right inset), it sacrifices energy efficiency, degrades spatial resolution, and still fails to fully eliminate the noise, ultimately deteriorating the overall performance (SSIM = 0.52). These results confirm that the DCL's multimode nature and nonlinear thresholding mechanism offer fundamental benefits compared to passive linear systems.

In the temporal domain, the inherent processing latency of the DCL is dictated solely by the $\sim$90~$\mu\mathrm{s}$ required for relaxation oscillations to settle into a steady state (Fig.~\ref{f5}D). This latency is a physical timescale governed by intrinsic cavity properties---such as the gain medium, cavity length, and pump level---and is independent of the input image size ($N\times N$). Consequently, the DCL demonstrates an $\mathcal{O}(1)$ computational complexity. This low latency enables real-time processing of dynamic inputs (see Supplementary Note 12 and Supplementary Movies). Currently, the overall throughput is practically limited by the SLM refresh interval ($\sim$0.11~$\mathrm{s}$).

We further quantitatively benchmark the DCL against standard digital image processing algorithms (detailed in Supplementary Note~13). For edge detection, we select classical operators including Sobel\cite{sobel1970camera}, Prewitt\cite{prewitt1970object}, Roberts\cite{roberts1963machine}, LoG\cite{marr1980theory}, and Canny\cite{canny1986computational}. For image denoising\cite{gonzalez2009digital}, the baselines include median filtering (MF), arithmetic mean filtering (AMF), low-pass filtering (LPF), area opening (AO), and morphological opening (MO). As demonstrated in Figs.~\ref{f5}E and F, our laser processor operates orders of magnitude faster than most digital counterparts (Figs.~\ref{f5}E(i) and F(i)), which suffer from a quadratic scaling $\mathcal{O}(N^2)$ complexity (Figs.~\ref{f5}E(ii) and F(ii)). Crucially, the joint visualization of SSIM and runtime in Fig.~\ref{f5}F(i) demonstrates the overall denoising superiority of the DCL. While some algorithms (i.e., AO and MO) yield higher SSIM values, a visual inspection reveals distinct structural erosion and erroneous feature deletion (detailed in Supplementary Note~13). These limitations are naturally circumvented by the gain-loss interplay within the DCL. Although more advanced digital approaches, such as neural networks\cite{xie2015HED, zhang2017DnCNN}, could achieve superior fidelity, they incur massive computational overhead and heavy dataset training costs. In contrast, the DCL processor is training-free.

\section*{Discussion}
\noindent{We} have demonstrated a physics-driven computational scheme for optical image processing, which transitions from conventional passive architectures to an active laser-based framework. By utilizing a multimode laser resonator as the core computing module, we reframe image processing as spontaneous mode selection governed by the intracavity gain-loss interplay. Programmable loss manipulation enables the system to autonomously achieve high-fidelity edge detection and robust denoising; switching between the two tasks simply requires updating the input encoding strategy. The synergy of resonant enhancement, parallel multimode oscillation, intrinsic nonlinearity, and ultrafast lasing dynamics endows our processor with distinct advantages when benchmarked against both passive optical filtering and standard digital algorithms. Specifically, it uniquely delivers considerable intracavity intensity for frequency upconversion, high-quality outputs free of coherent artifacts, as well as low latency with an $\mathcal{O}(1)$ complexity.

By further integrating Q-switching or longitudinal mode-locking techniques, the fundamental physical latency of the DCL could be dramatically compressed to the nanosecond or even femtosecond regime\cite{chriki2018spatiotemporal,mahler2020improved,liu2025generation}. To construct a truly all-optical computing architecture, future work could explore the incorporation of optically addressed modulators as input interfaces\cite{fan2026spatial}, thereby eliminating the latency and power consumption associated with optoelectronic conversions. Furthermore, the potential of our platform could extend to optical metrology and microscopy\cite{mader2015scanning,juffmann2016multi}. For instance, a transmissive or reflective specimen could be inserted into the cavity such that its microscopic features in thickness or absorption would dictate the loss distribution, thereby enabling highly sensitive, noise-suppressed phase or amplitude imaging. Beyond image processing, by leveraging the thresholding mechanism to physically implement a parallel ReLU-like nonlinear response across massive spatial modes, this platform emerges as a compelling candidate for the parallel nonlinear activation layer in future deep optical neural networks\cite{shi2025review}.

Broadly, this platform represents a deep convergence of laser physics and optical computing. By harnessing the complex lasing dynamics of a highly multimode cavity, we provide a scalable, fast, and reconfigurable solution for multifunctional image processing. Ultimately, this work establishes a promising hardware foundation for next-generation smart photonic sensors and all-optical neural networks.

\vspace{0.5cm}

\noindent\textbf{Methods}
\medskip
\begin{footnotesize}

\noindent\textbf{Experimental setup.} The detailed experimental setup of the DCL-based image processing system is described in Supplementary Note~1. The gain medium is an $a$-cut, 1~at.\%-doped Nd:YVO$_4$ crystal (diameter: 17~$\mathrm{mm}$; thickness: 2~$\mathrm{mm}$), with its rear surface high-reflection coated at 1064~nm to serve as one end mirror of the resonator. The other end mirror is a reflective phase-only SLM (X15213-03BL, Hamamatsu) with a pixel pitch of 12.5~$\mu\mathrm{m}$. The intracavity 4$f$ system is formed by two achromatic doublets with focal lengths of 500~$\mathrm{mm}$ and 150~$\mathrm{mm}$. An intracavity polarizing beam splitter (PBS) combined with a half-wave plate (HWP) is used to control the polarization state and couple a fraction of the oscillating field to a CMOS camera for detection. The system is pumped by a fiber-coupled laser diode centered at 878.6~$\mathrm{nm}$, operating in a quasi-continuous-wave (QCW) mode with a pulse width of 400~$\mu\mathrm{s}$ and a repetition rate of 10~$\mathrm{Hz}$. To investigate the stabilization latency of the DCL-based processor, we analyze the temporal evolution of the output intensity within a single 400~$\mu\mathrm{s}$ pump window. The output field is focused onto a high-speed photodetector (HDETIN04-20G-L, LBTEK), and the electrical signal is recorded using an oscilloscope (MSO73304DX, Tektronix).

\vspace{0.1cm}

\noindent\textbf{Simulation model.} To elucidate the underlying physical mechanism of intracavity mode evolution and to corroborate the experimental observations, we numerically model the DCL based on the Fox--Li algorithm\cite{fox1961resonant}. In a degenerate cavity, a single Fox--Li iteration converges to an initial-condition-dependent coherent superposition of many degenerate modes\cite{wolf1963spatial,wolf1984coherence}. To accurately model the experimentally measured intensity distribution, we incoherently average the steady-state outputs of multiple independent Fox--Li realizations. For each realization, the round-trip evolution of the cavity field is calculated by iteratively applying a sequence of mathematical operators representing physical diffraction, phase modulation, and aperture truncation. Leveraging the intracavity 4$f$ configuration, the diffractive propagation of the light field between the front and rear focal planes of the intracavity lenses is modeled as a Fourier transform operation. Additionally, to faithfully capture the experimental lasing dynamics, our model explicitly incorporates the nonlinear saturable gain of the gain medium. Comprehensive details regarding the simulation framework and physical parameters are provided in Supplementary Note~10.

The numerical model is implemented in MATLAB. To accelerate the simulations, the computationally intensive operations within the core iterative loops are offloaded to a GPU. All simulations are performed on a local workstation equipped with an Intel Core i5-10400F CPU and an NVIDIA GeForce RTX 3060 Ti GPU. Each complete simulation to generate a single output takes approximately 280~seconds, whereas the real optical system naturally achieves the result within $\sim$90~$\mu\mathrm{s}$.

\vspace{0.1cm}

\noindent\textbf{Image pre-processing and post-processing methods.} Detailed pre-processing steps applied to all input images for edge detection and image denoising are summarized in Supplementary Note~3. For the output data, region of interest (ROI) selection is applied to the full field-of-view raw images captured by the camera, as illustrated in Supplementary Note~4. To preserve the relative contrast information, each output image is normalized by the maximum intensity within the ROI. Notably, only ROI cropping and linear normalization are used, with no further computational adjustments applied.

\vspace{0.1cm}

\noindent\textbf{Nonlinear thresholding behavior of the laser.} For a given spatial mode, the relationship between output laser power $P_{\mathrm{out}}$ and optical pump power $P_{\mathrm{pump}}$  follows:
\begin{equation}
P_{\mathrm{out}}=
\begin{cases}
0, & P_{\mathrm{pump}}<P_{\mathrm{th}},\\
\eta_\mathrm{s}\left(P_{\mathrm{pump}}-P_{\mathrm{th}}\right), & P_{\mathrm{pump}}\ge P_{\mathrm{th}},
\end{cases}
\end{equation}
where the pump threshold $P_{\mathrm{th}}$ is proportional to the modal round-trip loss $\delta$ ($P_{\mathrm{th}}\propto \delta$) and the slope efficiency $\eta_\mathrm{s}$ scales inversely with $\delta$ ($\eta_\mathrm{s}\propto 1/\delta$). Therefore, spatial modes with different losses exhibit varying threshold pump powers and slope efficiencies. Under a uniform pump, different modes are selectively excited, and the corresponding thresholding behavior intrinsic to laser dynamics is the core nonlinear mechanism of this work in processing images.

\vspace{0.1cm}

\noindent\textbf{Evaluation of edge detection results.} To validate the edge detection capability of the DCL, one-dimensional cross-sectional intensity profiles, $I(x)$, are extracted along the red dashed lines indicated in each simulated and experimental output in Figs.~\ref{f2} and \ref{f3}. Since the image boundaries manifest as intensity depressions within the laser output, the edge locations are identified by detecting the local maxima of the inverted profile, $-I(x)$, using the \texttt{findpeaks} function in MATLAB. To establish a unified standard for evaluating the edge extraction capability across varying experimental configurations, the \texttt{MinPeakProminence} and \texttt{MinPeakHeight} parameters are fixed at 0.1 and -0.6, respectively. The edges identified by this method are marked with red circles in the corresponding intensity profiles. Notably, this local-valley-finding approach relies on relative intensity gradients rather than absolute values. It is therefore insensitive to global variations in the spot brightness, effectively eliminating the influence of the spatial inhomogeneity of the pump.

To quantitatively assess and compare the edge detection performance for the experimental results presented in Fig.~\ref{f3}, we calculate the edge visibility of the laser outputs. The visibility is quantified by the contrast metric $V=(I_{\max}-I_{\min})/(I_{\max}+I_{\min})$, where $I_{\max}$ and $I_{\min}$ denote the maximum and minimum intensities within the ROI, respectively. The calculated visibility value $V$ is displayed in the lower-left corner of each experimental result.

\vspace{0.1cm}

\noindent\textbf{Quantitative denoising quality assessment.} To quantitatively evaluate the denoising performance, we employ two standard metrics: the peak signal-to-noise ratio (PSNR) and the structural similarity index (SSIM). The PSNR measures the pixel-wise fidelity and is defined as
\begin{equation}
\mathrm{PSNR}=10\cdot \log_{10}\!\left(\frac{\mathrm{MAX}_\mathrm{I}^{2}}{\mathrm{MSE}}\right),
\end{equation}
where $\mathrm{MAX}_\mathrm{I}$ denotes the dynamic range of the pixel values (fixed at 1 for our normalized images), and $\mathrm{MSE}$ denotes the mean squared error between the denoised output ($I_{\mathrm{out}}$) and the noise-free ground truth ($I_{\mathrm{ref}}$). The SSIM provides a perceptual assessment of structural information preservation and is calculated as
\begin{equation}
\mathrm{SSIM}\!\left(I_{\mathrm{out}},I_{\mathrm{ref}}\right)=
\frac{\left(2\mu_{\mathrm{out}}\mu_{\mathrm{ref}}+C_{1}\right)\left(2\sigma_{\mathrm{out,ref}}+C_{2}\right)}
{\left(\mu_{\mathrm{out}}^{2}+\mu_{\mathrm{ref}}^{2}+C_{1}\right)\left(\sigma_{\mathrm{out}}^{2}+\sigma_{\mathrm{ref}}^{2}+C_{2}\right)},
\end{equation}
where $\mu_{\mathrm{out}}$ and $\mu_{\mathrm{ref}}$ denote the local means of $I_{\mathrm{out}}$ and $I_{\mathrm{ref}}$, respectively; $\sigma_{\mathrm{out}}^{2}$ and $\sigma_{\mathrm{ref}}^{2}$ are their corresponding local variances; and $\sigma_{\mathrm{out,ref}}$ represents the cross-covariance between the two images. Constants $C_{1}=0.01^{2}$ and $C_{2}=0.03^{2}$ are introduced to ensure stability. Higher PSNR and SSIM indicate better reconstruction quality.

\vspace{0.1cm}

\noindent \textbf{Acknowledgements}: This work was supported by the National Natural Science Foundation of China (62275137, 623B2064), the Fundamental and Interdisciplinary Disciplines Breakthrough Plan of the Ministry of Education of China (JYB2025XDXM121), and the Shenzhen Science and Technology Program (No. CJGJZD20240729141102004). \vspace{0.1cm}

\noindent \textbf{Author contributions}: J.W. and H.W. conceived the project. J.W. performed the experiments, simulations, and data analysis, with the help of H.W. and Y.Y.. J.W., H.W., and J.H. prepared the manuscript with inputs from all authors. X.F. and Q.L. supervised the research.
\vspace{0.1cm}

\noindent \textbf{Competing interests}: The authors declare they have no competing interests.
\end{footnotesize}

\newpage
\renewcommand{\bibpreamble}{
$^\ast${Corresponding authors: \textcolor{magenta}{hao.wang37@cityu.edu.hk}, \textcolor{magenta}{fuxing@tsinghua.edu.cn}, \textcolor{magenta}{qiangliu@tsinghua.edu.cn}}\\
}

\bibliographystyle{naturemag}
\bibliography{ref}

\end{document}